\documentclass[sigconf, pbalance=true]{acmart}

\setcopyright{none}
\acmDOI{}
\acmISBN{}
\acmConference[LOCO 2024]{1st International Workshop on Low Carbon Computing}{December 3, 2024}{Glasgow/Online}
\acmYear{2024}

\usepackage{graphicx} \usepackage[capitalise, nameinlink, sort]{cleveref}

\usepackage[]{todonotes}

\usepackage{xspace}
\newcommand{\eg}{e.g.\xspace}
\newcommand{\ie}{i.e.\xspace}
\makeatletter
\newcommand*{\etc}{\@ifnextchar{.}{etc}{etc.\@\xspace}}
\makeatother

\begin{document}

\title{A Digital Twinning Approach to Decarbonisation: Research Challenges}

\author{Blair Archibald}
\affiliation{\institution{University of Glasgow}
  \city{Glasgow}
  \country{Scotland}
}

\author{Paul Harvey}
\affiliation{\institution{University of Glasgow}
  \city{Glasgow}
  \country{Scotland}
}

\author{Michele Sevegnani} 
\affiliation{\institution{University of Glasgow}
  \city{Glasgow}
  \country{Scotland}
}

\maketitle

\section*{Introduction}
Transportation accounts for around 27\% of green house gas emissions in the UK~\cite{UKgov}. While an obvious priority area for decarbonisation, and aligned to the UK government goal of reducing emissions by 68\% for 2030, the free-market nature of the transportation sector combined with its fundamentally implicit and pervasive connections to all aspects of society and national infrastructure mean that all decarbonisation efforts to date have been siloed within a single transport sector~\cite{TransiT}, \eg only considering greener aviation fuels.
Truly decarbonising transport requires radical changes to the entire transport infrastructure, and since that transport does not happen in isolation, a single user often using multiple modes, we need a view over the whole transport system.

The first step to solving a problem is to understand it. As a result of the fragmented nature of the transportation sector, there is currently no system level view. Without the ability to monitor even adjacent transport domains, the ability for people or organisations to (dynamically) adapt their operations for decarbonisation outcomes is unrealistic. As transportation is a complex social-techno-economic system, information and knowledge sharing is a \emph{must} to be able to understand and explore potential solutions to the decarbonisation challenge.

We believe a \textbf{Federated Digital Twinning Approach} has the potential to tackle transport decarbonisation problems, and,
in this extended abstract, \textbf{we give an overview of the research required to tackle the fundamental challenges around digital twin design, generation, validation and verification.}

\section*{Federated Digital Twinning for Decarbonisation}
Digital Twins~\cite{semeraro2021digital} (DTs) are data-driven virtual representations of physical (or conceptual) assets. By applying transportation data to DTs, they can
provide a pathway to understanding the current transport system, and enable experimentation with future transport systems: a necessary ability for decarbonisation given that physical experimentation at scale is not feasible or timely.

\begin{figure}
    \centering
    \includegraphics[width=1.0\linewidth]{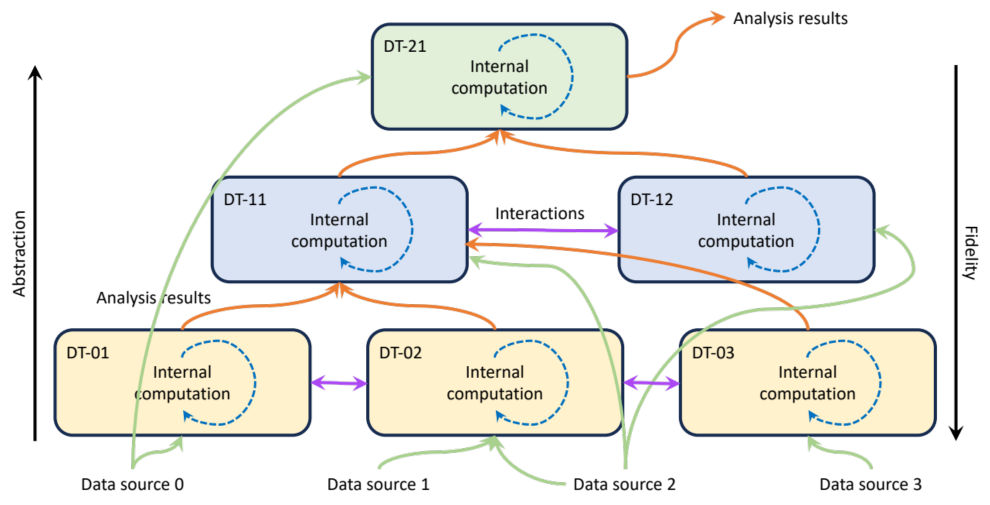}
    \caption{Federated Digital Twinning Approach}
    \label{fig:federation}
\end{figure}

DTs may take many forms: from simple displays of data (dashboards), machine learning-based data labelling (convolutional neural networks), large-scale simulations (discrete event simulators), numerical models based on partial differential equations (weather prediction), 3D models, and logical models building on discrete mathematical representations of entities and their interactions. 
While many useful mode-specific closed-loop DTs exist within transport, \eg Port of Dover Twin~\cite{Dover} and elsewhere~\cite{9795043}, they remain siloed, lacking interoperability or a standardised framework for communication. For example, the port only considers port operations, excluding freight shipping entering and road congestion leaving the port. 
To overcome the lack of information sharing in the transport sector we propose \textit{federated digital twinning approach for transport decarbonisation} as shown in \cref{fig:federation}.

Federated twins compose multiple DTs together to give better operational oversight.
This is a step change from classic digital twins that only have interoperability between the physical system and a single twin.
Moving to a federated twinning approaches raises many interesting research challenges.

\section*{Federated Digital Twinning Research Challenges}

We seek to derive value from the federation of multiple twins, possibly owned by different stakeholders, \eg utilising power generation models alongside traffic and road models to develop effective decarbonisation policies.
Fundamentally, integrating Digital Twins means we need \textbf{approaches to combine their underlying models} and to be sure these combinations make logical sense. This is a challenge: these models are heterogeneous and we need a well-defined and common theory to be able to use them together. 
While domain-specific modelling tools have been extensively developed over the years~\cite{Azure, Omniverse, Bentley}, the techniques for combining models are under-explored but essential. 

This combined modelling approach is further complicated by: 1.~\textbf{multi-scale modelling}---one twin might be for a single component, \eg a car, while another may be at a completely different level of abstraction or fidelity, \eg city-scale; 2.~\textbf{multi-temporal modelling}---models work at different time scales and some modelling techniques will trade granularity for performance, \eg it is quicker to make a weather prediction based on seasons than a full atmospheric weather simulation; 3.~\textbf{ integration of multiple model types}, e.g. discrete models of software behaviours, continuous models of physical behaviours, and probabilistic models of environment, devices failure rates, and data accuracy; and 4.~\textbf{dynamic models}---the systems we model are subject to change, \eg deployment of more 5G/6G cells, new road infrastructure being built, emergent behaviour of crowds, and these need to be integrated into the models. 
Throughout all these challenges there is a need for rigour: transport systems are critical infrastructure, have dramatic effects on peoples' lives, and we do not have time to take multiple-shots at getting the decarbonisation right.

\subsection*{Proposed Approach}

\textbf{We propose to develop new theories and tools to reason about federated DT models based on Formal Methods}—mathematical descriptions of systems that allow for strong reasoning, \eg creating proofs of correctness via mathematical logic. Importantly, the model specification will be driven by the incoming data\footnote{Which is a defining feature of digital twins versus traditional modelling approaches.}, but not fully defined by it. In this way we maintain control of core reasoning aspects, \ie badly inferred models can be disregarded by a human-in-the-loop. 
Newer formal modelling approaches allow for describing the probabilistic scenarios that are essential when working with real data that sometimes goes missing, is incorrect or non-trustworthy; and for describing the spatial elements of systems, which is crucial when working with transport and other physical systems.

Formal theories will be applied when reasoning about systems themselves, and, at a meta-level to reason about the integration of Digital Twins. 
For example, we may model dataflow between twins to detect possible attack vectors such as a malicious twin using electric vehicle power data to detect areas of stress on the power grid and other critical infrastructure. 
Bigraphs~\cite{DBLP:books/daglib/0022395,DBLP:journals/tcs/SevegnaniC15} may form the common (formal) meta-model for describing federated Digital Twins.
Bigraphs model systems with strong notions of spatial locality, \eg a car within a particular charging station, and non-local connectivity, \eg modelling the communication between smart traffic lights and a central control room.
A key benefit of bigraphs is the visual modelling format---not too unlike what you might draw on a whiteboard---making them accessible to a wide range of transport experts without needing detailed formal modelling knowledge.

For analysis, bigraphs explicitly represent \textbf{all} possible temporal evolutions of a system as opposed to simulation which only considers a set of traces. Bigraphs support  underlying stochastic process (\eg Markov Chains and Markov Decision Processes) describing the evolution of a system in terms of probabilities or rates, giving access to automated verification with tools like PRISM~\cite{DBLP:conf/cav/KwiatkowskaNP11} to answer queries such as ``given a battery failure probability of 0.05, is the probability of running out of charge before reaching a charger less than 0.001?''. By changing the model (\ie bigraph structure, reaction rules, or probabilities) and re-running the query we can perform \emph{what-if} scenarios.

A single model will not be enough for all situations, and there are interesting questions around whether we want models to prove slow, but accurate, or fast, with wider confidence interval results.
Partially Observable Markov Decision Processes (POMDPs) will be well-suited for our requirements as they encompass uncertainty on the state observation. However, the theory of bigraphs will require extensions as it does not yet support POMDPs. Another important feature of POMDPs is that they support the synthesis of strategies (\ie policies), that allow the models to perform optimisation with respect to metrics like decarbonisation.
We also focus on how to parametrise Digital Twins models by the data stream generated and transmitted by the physical system, \ie we need live models@runtime~\cite{DBLP:journals/computer/BlairBF09, DBLP:conf/iceccs/SevegnaniKCM18}. We expect out-of-order messages and mislabelling to be potential issues, but these fit those as another form of uncertainty in the model. 

When designing and managing large-scale systems involving multiple agents and stakeholders it is challenging to reason about (future) properties of the overall system \eg CO$_2$ emissions over time. We will express these kinds of properties using rewards in Continuous Stochastic Logic (PCTL)~\cite{DBLP:journals/fac/HanssonJ94}, and we will verify them through model checking using PRISM. 
An additional challenge is enabling non-expert users to express these properties, and we believe approaches like FRET~\cite{DBLP:conf/refsq/FarrellLSM22} and nl2spec~\cite{DBLP:conf/cav/CoslerHMST23} to be of use here.

We will develop extensions of the standard verification algorithms to support checking a set of properties across a hierarchy of Digital Twins and not just over one. For models with large state spaces, exhaustive model checking can be computationally expensive, so we will utilise the built-in discrete-event simulator in PRISM to apply statistical model checking (SMC)~\cite{DBLP:conf/cav/YounesS02} and compute approximately correct results. SMC effectively samples the model space through repeated simulation instead of exhaustive search resulting in higher performance at a cost of accuracy. Building on recent advances in the field~\cite{DBLP:journals/corr/CianciaLLM16,DBLP:conf/mfcs/LinkerPS21,DBLP:conf/pads/KreikemeyerHU20}, we will define more refined properties using spatial logics.

Open questions remain around how the end-users will specify the models: while the visual approach of bigraphs can be exploited, it is possible we require additional programming language support: what does a programming language specifically for digital twin models look like? 
A complementary approach is mechanically generating suitable models of each component of the system via process mining or generative AI techniques.
This requires discretisation and different levels of abstraction and formulate precise relations between models to form an overall coherent hierarchy. 
Once we have the model running, how best to express the structure of the system to the user? We suggest a control interface updated with the results of model checking as the become available.

\section*{Conclusion}

A federated digital twinning approach has the potential to give the whole-system view required for timely and radical transport decarbonisation and missing in current approaches.

To fully utilise federated twins we must overcome challenges related to model combination across scales, fidelity, and modelling types.
We believe formal methods is a core component to solving this challenges. If we get it right, it provides the underlying trust we need for federated twins of such critical importance.

\begin{acks}
This work is supported by the Engineering and Physical Sciences Research Council, under grant EP/Z533221/1 (TransiT: Digital Twinning Research Hub for Decarbonising Transport) and an Amazon Research Award on Automated Reasoning.
\end{acks}

\bibliographystyle{ACM-Reference-Format}

\end{document}